\documentclass{Interspeech}
\usepackage{times}
\usepackage{latexsym}
\usepackage{booktabs}
\usepackage{verbatim}
\usepackage[table,xcdraw]{xcolor}
\usepackage{multirow}
\usepackage[table,xcdraw]{xcolor}
\usepackage{colortbl}  
\usepackage[table,xcdraw]{xcolor}
\usepackage{booktabs}
\usepackage{hyperref}
\usepackage[T1]{fontenc}

\usepackage{etoolbox}
\makeatletter
\patchcmd{\affiliation}
  {\renewcommand{\affiliationsep}{\vskip1pt}}
  {\renewcommand{\affiliationsep}{, }}
  {}{}
\makeatother

\usepackage[utf8]{inputenc}
\usepackage{microtype}
\definecolor{lightred}{rgb}{1.0, 0.8, 0.8}
\definecolor{lightblue}{rgb}{0.8, 0.9, 1.0}
\definecolor{lightgreen}{rgb}{0.8, 1.0, 0.8}
\definecolor{lightyellow}{rgb}{1.0, 1.0, 0.8}
\definecolor{lightpurple}{rgb}{0.9, 0.8, 1.0}
\definecolor{lightorange}{rgb}{1.0, 0.9, 0.8}
\usepackage{inconsolata}
\usepackage{adjustbox}
\usepackage{ subcaption, algorithm2e, url, float, etoolbox, adjustbox, pgf,booktabs, verbatim}
\usepackage{xcolor}
\usepackage{amsmath}
\usepackage{amssymb}
\usepackage{tcolorbox}

\definecolor{mattePink5}{RGB}{255,240,245}    
\definecolor{mattePink10}{RGB}{255,225,235}
\definecolor{mattePink15}{RGB}{255,210,225}
\definecolor{mattePink20}{RGB}{255,195,215}
\definecolor{mattePink25}{RGB}{255,180,205}    

\title{Investigating the Reasonable Effectiveness of Speaker Pre-Trained Models and their Synergistic Power for SingMOS Prediction}
\author[affiliation={1}]{Orchid Chetia}{Phukan*}
\author[affiliation={1,2}]{Girish*}{}
\author[affiliation={1,3}]{Mohd Mujtaba}{Akhtar*}
\author[affiliation={4}]{Swarup Ranjan}{Behera}
\author[affiliation={5}]{Pailla Balakrishna}{Reddy}
\author[affiliation={1}]{Arun Balaji}{Buduru}
\author[affiliation={6,7}]{Rajesh}{Sharma}

\affiliation{}{IIIT-Delhi}{India}
\affiliation{}{UPES}{India}
\affiliation{}{V.B.S.P.U}{India}
\affiliation{}{Independent Researcher}{India}
\affiliation{}{Reliance AI}{India}
\affiliation{}{University of Tartu}{Estonia}
\affiliation{}{Plaksha University}{India}
\email{\textcolor{blue}{\texttt{Correspondence:}} orchidp@iiitd.ac.in} 

\usepackage{comment}

\begin{document}

\maketitle
\begingroup
  \renewcommand{\thefootnote}{\fnsymbol{footnote}}
  \setcounter{footnote}{0}
   \footnotetext{* Contributed equally as a first authors.}
\endgroup

\begin{abstract}

\noindent In this study, we focus on Singing Voice Mean Opinion Score (SingMOS) prediction. Previous research have shown the performance benefit with the use of state-of-the-art (SOTA) pre-trained models (PTMs). However, they haven't explored speaker recognition speech PTMs (SPTMs) such as x-vector, ECAPA and we hypothesize that it will be the most effective for SingMOS prediction. We believe that due to their speaker recognition pre-training, it equips them to capture fine-grained vocal features (e.g., pitch, tone, intensity) from synthesized singing voices in a much more better way than other PTMs. Our experiments with SOTA PTMs including SPTMs and music PTMs validates the hypothesis. Additionally, we introduce a novel fusion framework, \texttt{\textbf{BATCH}} that uses Bhattacharya Distance for fusion of PTMs. Through \texttt{\textbf{BATCH}} with the fusion of speaker recognition SPTMs, we report the topmost performance comparison to all the individual PTMs and baseline fusion techniques as well as setting SOTA.


\end{abstract}

\section{Introduction}
Mean Opinion Score (MOS) is a foundational metric for evaluating audio quality, widely employed across speech and singing domains. Automatic MOS prediction offers numerous benefits, including efficiency and scalability, as it eliminates the need for time-intensive human evaluations and enables rapid assessment of large-scale samples generated through generative systems such as TTS (text-to-speech) and VC (voice conversion). It is also highly cost-effective, reducing the reliance on human annotators, and ensures consistency and objectivity by providing uniform and unbiased evaluations. Speech MOS prediction have seen significant advancements due to the availability of labelled datasets such as SOMOS \cite{maniati22_interspeech} and VoiceMOS \cite{cooper2023voicemos}. As such researchers have explored various techniques for speech MOS prediction \cite{qi2023ssl, vioni2023investigating} and the current research scenario is mostly dependent on the use of state-of-the-art (SOTA) pre-trained models (PTMs) trained on large-scale diverse datasets \cite{kunevsova2023ensemble, 10688047, linnon, udupa24b_interspeech}. 

Despite progress in MOS prediction systems, the task of singing voice MOS (SingMOS) prediction has remained relatively underexplored compared to speech MOS prediction systems. The demand for automated singing quality assessment in applications such as music production and virtual performances underscores the importance of SingMOS prediction. Tang et al. \cite{tang2024singmos} made the first contribution towards SingMOS prediction by introducing the first SingMOS prediction dataset, followed by a broader exploration of speech PTMs (SPTMs) for this task \cite{tang2024exploration}. Here, they have explored various SOTA SPTMs such XLS-R, wav2vec2, and so on that includes both monolingual and multilingual SPTMs. Additionally, Huang et al. \cite{huang2024mos} also explored wav2vec2 for SingMOS prediction. However, previous research haven't included speaker recognition SPTMs such as x-vector, ECAPA that are pre-trained primarily for recognizing speakers and which we believe will be the most effective for SingMOS prediction. \par

In this study, we investigate speaker recognition SPTMs for the first time for SingMOS prediction and \textit{hypothesize that speaker recognition SPTMs can significantly enhance SingMOS prediction due to their speaker recognition pre-training. This pre-training equips them with the capability to capture fine-grained attributes such as pitch, tone, intensity, rhythm - from singing voice most effectively which are the key elements in assessing synthesized singing voice quality.} To validate our hypothesis, we present a large-scale comparative study of speaker recognition, multilingual, monolingual SPTMs, and music PTMs (MPTMs). We incorporate MPTMs in SingMOS prediction due to their ability to model the melodic and tonal qualities inherent in singing voices, which are crucial for perceptual quality assessment. Our experiments on benchmark SingMOS prediction dataset through FCN (Fully Connected Network) and CNN downstreams validates our hypothesis. Additionally, motivated by research on speech MOS prediction \cite{yang22o_interspeech} as well as other speech processing tasks such as synthetic speech detection \cite{chetia-phukan-etal-2024-heterogeneity}, speech recognition \cite{arunkumar22b_interspeech} where researchers have shown the fusion of PTMs leads to improve performance, we also explore the same for SingMOS prediction. To this end, we present a novel framework, \texttt{\textbf{BATCH}} (Fusion via \texttt{\textbf{B}}h\texttt{\textbf{AT}}ta\texttt{\textbf{CH}}aryya Distance), which employs Bhattacharyya Distance as a novel loss function to fuse PTMs effectively. To the best of our knowledge, we are the first study to explore fusion of PTMs for improved SingMOS prediction. By aligning speaker recognition PTMs (x-vector and ECAPA) through \texttt{\textbf{BATCH}}, we report the topmost performance in comparison to all the individual PTMs and baseline fusion techniques. We also report SOTA in comparison to previous SOTA works. 

\noindent \textbf{To summarize, we make the following key contributions in this work:}
\begin{itemize}  
    \item We present a comprehensive evaluation of SOTA PTMs for SingMOS prediction, examining both SPTMs and MPTMs and demonstrating their efficacy in capturing the complex characteristics of singing quality. Our findings highlight SPTMs trained for speaker recognition as the top performers for SingMOS prediction.  
    \item We analyze the superior performance of x-vector (speaker recognition SPTM), attributing its success to speaker recognition pre-training, which enables it to effectively capture key vocal features - such as pitch, tone, and intensity - critical to accurate SingMOS prediction.  
    \item We introduce a novel framework, \texttt{\textbf{BATCH}} for the fusion of PTMs. With \texttt{\textbf{BATCH}}, through the fusion of x-vector and ECAPA (Both are speaker recognition SPTMs), we report SOTA performance in SingMOS prediction, surpassing previous benchmarks and revealing new insights for future research.
\end{itemize}
\noindent The full codebase and associated models will be made publicly available at: \url{https://github.com/Helix-IIIT-Delhi/BATCH-SingMOS}

\begin{figure}[!bt]
    \centering
    \includegraphics[width=0.9\linewidth]{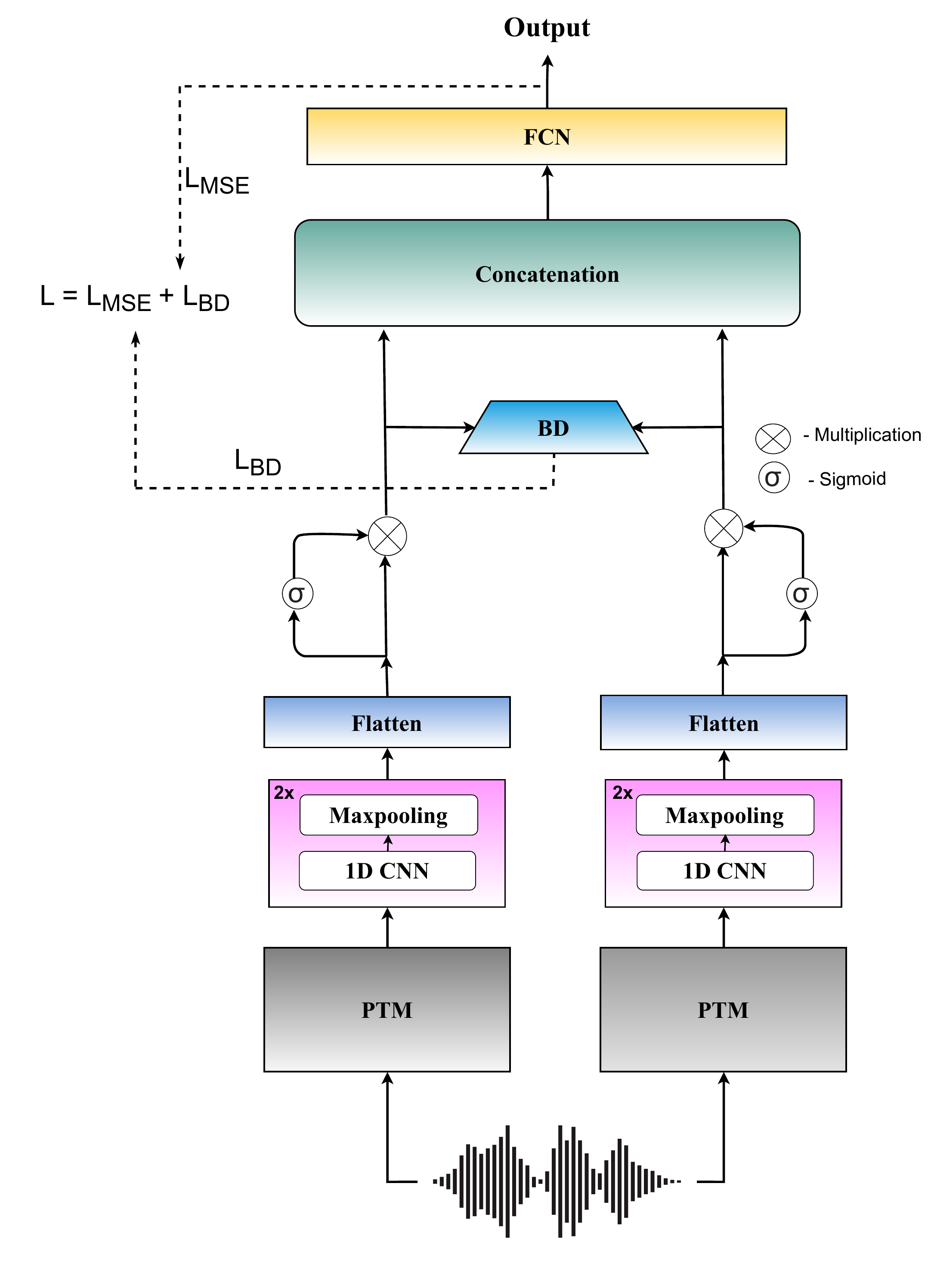}
    \caption{Proposed Framework: \texttt{\textbf{BATCH}}}
    \label{fig:archi}
\end{figure}

\section{Pre-Trained Models}

\noindent In this section, we explain the PTMs under consideration in our study. \par
\noindent\textbf{Speech PTMs}: We use x-vector\footnote{\url{https://huggingface.co/speechbrain/spkrec-xvect-voxceleb}} \cite{8461375} and ECAPA\footnote{\url{https://huggingface.co/speechbrain/spkrec-ecapa-voxceleb}} \cite{desplanques20_interspeech} as speaker recognition SPTMs. Both are time delay neural network and trained on combination of Voxceleb1 + Voxceleb2. ECAPA improves over x-vector whereas x-vector improves over i-vector for speaker recognition performance. x-vector contains 4.2M parameters approximately. As monolingual SPTMs, we leverage WavLM \cite{chen2022wavlm}, Unispeech-SAT\footnote{\url{https://huggingface.co/microsoft/unispeech-sat-base}} \cite{chen2022unispeech}, and Wav2vec2\footnote{\url{https://huggingface.co/facebook/wav2vec2-base}} \cite{baevski2020wav2vec}. We consider their base versions trained on librispeech 960 hours english data. WavLM and Unispeech-SAT both are SOTA SPTMs on diverse speech processing tasks within SUPERB. Unispeech-SAT is trained in a multi-task format whereas WavLM solves speech denoising and masked prediction simultaneously. Wav2vec2 was pre-trained to solve a constrastive task. We use WavLM, Unispeech-SAT, HuBERT, Wav2vec2 of 94.70M, 94.68M, and 95.04M parameters respectively. For multilingual SPTMs, we use XLS-R\footnote{\url{https://huggingface.co/facebook/wav2vec2-xls-r-300m}} \cite{babu22_interspeech}, Whisper\footnote{\url{https://huggingface.co/openai/whisper-base}} \cite{radford2023robust}, and MMS\footnote{\url{https://huggingface.co/facebook/mms-1b}} \cite{pratap2024scaling}. XLS-R, Whisper, MMS are pre-trained on 128, 96 and 1400 languages. XLS-R and MMS is built on top of Wav2vec2 architecture while Whisper follows a vanilla transformer-decoder architecture. We use 300M, 74M and 1B parameters versions of XLS-R, Whisper, and MMS respectively. We resample all the audios to 16KHz before passing it to the SPTMs and we extract representations from the last hidden state of the SPTMs by pooling average. For Whisper, we take out representations from the encoder. We get representations of following dimensions: 768: Unispeech-SAT, Wav2vec2, WavLM; 512: x-vector, Whisper; 192: ECAPA and 1280: MMS, XLS-R. 

\noindent\textbf{Music PTMs}: We use music2vec-v1\footnote{\url{https://huggingface.co/m-a-p/music2vec-v1}} \cite{Music2VecAS} and MERT-series of MPTMs \cite{li2023mert} for our experiments. We consider various MERT variants: MERT-v0\footnote{\url{https://huggingface.co/m-a-p/MERT-v0}}, MERT-v0-public\footnote{\url{https://huggingface.co/m-a-p/MERT-v0-public}}, MERT-v1-95M\footnote{\url{https://huggingface.co/m-a-p/MERT-v1-95M}}, and MERT-v1-330M\footnote{\url{https://huggingface.co/m-a-p/MERT-v1-330M}}. music2vec-v1 is a self-supervised learning MPTM pre-trained for usage in various music-related downstream tasks. MERT-series represents SOTA MPTMs across music processing tasks such music genre prediction as well as mood recognition. All the MPTMs except MERT-v1-330M is of 95M parameters which itself is of 330M parameters. All the MPTMs operate at certain audio sampling rates: MERT-v1-330M and MERT-v1-95M process audio at 24 kHz, while MERT-v0-public, MERT-v0, and music2vec-v1 use 16 kHz. Before processing, audio is resampled accordingly, and feature representations are extracted from frozen models using average pooling on the final hidden layer. The representation dimensions are 768 for music2vec-v1 and MERT versions, while MERT-v1-330M has a higher dimension of 1024.

\section{Modeling}
In this section, we discuss the downstream modeling with the PTMs followed by the proposed framework, \texttt{\textbf{BATCH}} for combination of PTMs.

\subsection{Individual Representation Modeling}

We use Fully Connected Network (FCN) and CNN as backend with individual PTMs. CNN comprises 1D convolutional layers with 64 and 128 filters, respectively with kernel size of 3 in both the layers. We apply maxpooling after each consecutive 1D convolutional layer. The output is flattened and routed through a FCN network that contains a dense layer with 128 neurons and ReLU activation followed by a output neuron for regression task. For FCN, we keep the same modeling details as the FCN used in CNN. 

\subsection{\texttt{BATCH}}

We propose a novel modeling framework, \texttt{\textbf{BATCH}} for the fusion of PTMs. The architecture is given in Figure \ref{fig:archi}. The PTM representations are first passed through two 1D convolutional layers and consecutive maxpooling layer as done with individual PTMs above. We keep the rest modeling same. Then the features are subjected to a gating mechanism. It is applied to dynamically modulate the extracted features. The output of the gated mechanism is computed as follows: $G(x) = \sigma(x) \odot x$, where $x$ is the input feature representation, $\sigma(\cdot)$ denotes the sigmoid activation function, and $\odot$ represents element-wise multiplication. This gating mechanism allows the model to selectively retain salient features while suppressing irrelevant ones. To align the PTMs feature representations, we introduce Bhattacharya Distance (BD) as a novel loss function. A higher BD indicates greater dissimilarity between the two distributions and we aim to optimize it for aligning the PTMs to a joint feature space. BD is defined as: $ D_B(P, Q) = -\log \left( \sum_x \sqrt{P(x) Q(x)} \right) $, where $P(x)$ and $Q(x)$ represent the feature distributions of the two PTMs at position $x$. The summation iterates over all positions in these feature representations, ensuring that the BD captures the overall similarity between the two PTMs representational space by comparing corresponding elements across their distributions. Unlike Euclidean distance, BD measures point-wise differences. This ensures better correlation between the two PTMs representations. It enhances robustness to variations, as traditional similarity measures like cosine similarity or KL divergence may struggle with heterogeneous statistical properties, whereas BD offers a more stable and reliable similarity measure. Additionally, BD encourages feature compactness by penalizing distant representations while promoting alignment, leading to improved generalization. After BD is calculated between the two PTMs feature space, the feature spaces are concatenated and passed through a FCN with 128 neurons followed by a output that predicts a continous value for SingMOS prediction. The BD loss value is then added to the mean squared error (MSE) loss for joint optimization. The total loss function is given by: $ L_{\text{total}} = L_{\text{MSE}} + \alpha L_{\text{BD}} $, where $L_{\text{MSE}} = \text{MSE Loss}$, $L_{\text{BD}}$ is the BD loss, and $\alpha$ is a tunable weight factor for controlling the contribution of the BD loss. The trainable parameters for \textbf{\texttt{BATCH}} ranges from 2M to 6M parameters depending on the input representation dimension size.

\section{Experiments}
\subsection{Dataset}

\vspace{-0.2cm}
We utilize the sole SingMOS dataset by Tang et al. \cite{tang2024singmos}. It comprises of 3,421 singing voice clips in Chinese and Japanese, sampled at 16 kHz, with a total duration of approximately 4.25 hours. Each clip has an average length of 4.47 seconds. The dataset includes a diverse range of audio samples generated by 21 singing voice synthesis (SVS) models, 6 singing voice conversion (SVC) systems, and 6 vocoder models. It is divided into three subsets: training, development, and testing. The test set is further split into two categories: test-main and test-other1. We train our models on the official train split, validate it on the development set and evaluate it on both the test sets. \newline
\noindent \textbf{Training Details}: We use Adam as the optimizer and MSE as the loss function. We train all the models with a batch size of 32 and for 50 epochs. We set the learning rate to 1e-3. Early stopping and dropout are employed to prevent overfitting. For experiments with \textbf{\texttt{BATCH}}, we set the value of $\alpha$ to 0.3 after initial experimentation. \newline

\begin{table}[!bt]
\setlength{\tabcolsep}{2pt}
\centering
\begin{tabular}{l|ll|ll|ll|ll}
\toprule
\multirow{2}{*}{} & \multicolumn{4}{c|}{\textbf{Test-other1}}                   & \multicolumn{4}{c}{\textbf{Test-main}}                     \\
\cmidrule(lr){2-5} \cmidrule(lr){6-9}
                  & \multicolumn{2}{c|}{\textbf{FCN}} & \multicolumn{2}{c|}{\textbf{CNN}} & \multicolumn{2}{c|}{\textbf{FCN}} & \multicolumn{2}{c}{\textbf{CNN}} \\
\cmidrule(lr){2-3} \cmidrule(lr){4-5} \cmidrule(lr){6-7} \cmidrule(lr){8-9}
                  & \textbf{MAE}   & \textbf{MSE}   & \textbf{MAE}   & \textbf{MSE}   & \textbf{MAE}   & \textbf{MSE}   & \textbf{MAE}   & \textbf{MSE}   \\
\midrule
U    & \cellcolor{mattePink5}1.00   & \cellcolor{mattePink5}1.35   & \cellcolor{mattePink20}0.79   & \cellcolor{mattePink20}0.90   & \cellcolor{mattePink20}0.73   & \cellcolor{mattePink25}0.79   & \cellcolor{mattePink15}0.45   & \cellcolor{mattePink15}0.53 \\
W2   & \cellcolor{mattePink15}0.86   & \cellcolor{mattePink20}1.05   & \cellcolor{mattePink20}0.77   & \cellcolor{mattePink15}0.97   & \cellcolor{mattePink15}0.85   & \cellcolor{mattePink20}1.00   & \cellcolor{mattePink15}0.46   & \cellcolor{mattePink10}0.55 \\
W    & \cellcolor{mattePink15}0.89   & \cellcolor{mattePink20}1.10   & \cellcolor{mattePink10}0.86   & \cellcolor{mattePink10}1.04   & \cellcolor{mattePink25}0.50   & \cellcolor{mattePink25}0.56   & \cellcolor{mattePink10}0.49   & \cellcolor{mattePink10}0.56 \\
X    & \cellcolor{mattePink15}0.90   & \cellcolor{mattePink5}1.42    & \cellcolor{mattePink5}0.89    & \cellcolor{mattePink5}1.11    & \cellcolor{mattePink10}1.21   & \cellcolor{mattePink10}1.83   & \cellcolor{mattePink15}0.43   & \cellcolor{mattePink15}0.53 \\
Wh   & \cellcolor{mattePink5}1.00    & \cellcolor{mattePink5}1.35    & \cellcolor{mattePink20}0.79   & \cellcolor{mattePink15}1.01   & \cellcolor{mattePink20}0.81   & \cellcolor{mattePink20}0.94   & \cellcolor{mattePink20}0.42   & \cellcolor{mattePink15}0.52 \\
M    & \cellcolor{mattePink10}0.95   & \cellcolor{mattePink10}1.28   & \cellcolor{mattePink15}0.84   & \cellcolor{mattePink5}1.19    & \cellcolor{mattePink10}1.13   & \cellcolor{mattePink15}1.60   & \cellcolor{mattePink10}0.49   & \cellcolor{mattePink15}0.51 \\
\textbf{XV}   & \cellcolor{mattePink25}\textbf{0.76}   & \cellcolor{mattePink25}\textbf{0.88}   & \cellcolor{mattePink25}\textbf{0.71}   & \cellcolor{mattePink25}\textbf{0.75}   & \cellcolor{mattePink25}\textbf{0.41}   & \cellcolor{mattePink25}\textbf{0.44}   & \cellcolor{mattePink25}\textbf{0.37}   & \cellcolor{mattePink25}\textbf{0.39} \\
\textbf{EC}   & \cellcolor{mattePink25}\textbf{0.78}   & \cellcolor{mattePink25}\textbf{0.88}   & \cellcolor{mattePink25}\textbf{0.72}   & \cellcolor{mattePink25}\textbf{0.76}   & \cellcolor{mattePink25}\textbf{0.42}   & \cellcolor{mattePink25}\textbf{0.40}   & \cellcolor{mattePink25}\textbf{0.31}   & \cellcolor{mattePink25}\textbf{0.39} \\
m2v  & \cellcolor{mattePink10}0.91   & \cellcolor{mattePink15}1.13   & \cellcolor{mattePink10}0.87   & \cellcolor{mattePink10}1.08   & \cellcolor{mattePink25}0.60   & \cellcolor{mattePink20}0.62   & \cellcolor{mattePink5}0.60    & \cellcolor{mattePink5}0.61 \\
MT95 & \cellcolor{mattePink10}0.95   & \cellcolor{mattePink10}1.24   & \cellcolor{mattePink5}0.93    & \cellcolor{mattePink5}1.18    & \cellcolor{mattePink20}0.62   & \cellcolor{mattePink20}0.65   & \cellcolor{mattePink5}0.57    & \cellcolor{mattePink5}0.63 \\
MTP  & \cellcolor{mattePink10}0.91   & \cellcolor{mattePink20}1.10   & \cellcolor{mattePink5}0.90    & \cellcolor{mattePink10}1.08   & \cellcolor{mattePink20}0.62   & \cellcolor{mattePink20}0.64   & \cellcolor{mattePink5}0.58    & \cellcolor{mattePink5}0.62 \\
MT3M & \cellcolor{mattePink10}0.92   & \cellcolor{mattePink5}1.45    & \cellcolor{mattePink10}0.87   & \cellcolor{mattePink10}1.04   & \cellcolor{mattePink5}1.42    & \cellcolor{mattePink5}2.53    & \cellcolor{mattePink5}0.58    & \cellcolor{mattePink5}0.62 \\
MTV0 & \cellcolor{mattePink10}0.93   & \cellcolor{mattePink15}1.16   & \cellcolor{mattePink5}0.90    & \cellcolor{mattePink15}1.13   & \cellcolor{mattePink20}0.62   & \cellcolor{mattePink15}1.63   & \cellcolor{mattePink5}0.61    & \cellcolor{mattePink5}0.63 \\
\bottomrule
\end{tabular}
\caption{Evaluation scores of various PTMs with FCN and CNN downstream networks; For each metric (MAE, MSE), lower values are better. Hence the lowest (best) values are shaded darkest and the highest (worst) values lightest; Abbreviations: U (Unispeech-SAT), W2 (Wav2vec2), W (WavLM), Hu (HuBERT), X (XLS-R), Wh (Whisper), M (MMS), XV (x-vector), EC (ECAPA), m2v (music2vec-v1), MT95 (MERT-v1-95M), MTP (MERT-v0-public), MT3M (MERT-v1-330M), MTV0 (MERT-v0); These abbreviations and color shades used in this Table are kept same for Table \ref{tab:fusion}}
\label{tab:single}
\end{table}

\begin{table}[]
\setlength{\tabcolsep}{3pt}
\scriptsize
\begin{tabular}{l|cc|cc|cc|cc}
\toprule
\multicolumn{1}{c|}{\multirow{3}{*}{\textbf{Combination}}} & \multicolumn{4}{c|}{\textbf{Concatenation}} & \multicolumn{4}{c}{\textbf{BATCH}} \\ \cmidrule{2-9}
 & \multicolumn{2}{c|}{\textbf{TO1}} & \multicolumn{2}{c|}{\textbf{TM}} & \multicolumn{2}{c|}{\textbf{TO1}} & \multicolumn{2}{c}{\textbf{TM}} \\ \cmidrule{2-9}
 & \textbf{MAE} & \textbf{MSE} & \textbf{MAE} & \textbf{MSE} & \textbf{MAE} & \textbf{MSE} & \textbf{MAE} & \textbf{MSE} \\ \midrule
U+W2    & \cellcolor{mattePink10}0.97 & \cellcolor{mattePink15}1.27 & \cellcolor{mattePink25}0.40 & \cellcolor{mattePink15}0.50 & \cellcolor{mattePink15}0.91 & \cellcolor{mattePink15}1.09 & \cellcolor{mattePink15}0.32 & \cellcolor{mattePink15}0.41 \\
U+W     & \cellcolor{mattePink10}1.02 & \cellcolor{mattePink10}1.37 & \cellcolor{mattePink15}0.41 & \cellcolor{mattePink15}0.51 & \cellcolor{mattePink15}0.90 & \cellcolor{mattePink15}1.04 & \cellcolor{mattePink15}0.32 & \cellcolor{mattePink15}0.42 \\
U+X     & \cellcolor{mattePink10}0.97 & \cellcolor{mattePink15}1.30 & \cellcolor{mattePink25}0.38 & \cellcolor{mattePink15}0.49 & \cellcolor{mattePink20}0.81 & \cellcolor{mattePink15}1.13 & \cellcolor{mattePink20}0.30 & \cellcolor{mattePink15}0.39 \\
U+Wh    & \cellcolor{mattePink10}1.01 & \cellcolor{mattePink10}1.38 & \cellcolor{mattePink15}0.41 & \cellcolor{mattePink15}0.51 & \cellcolor{mattePink20}0.79 & \cellcolor{mattePink20}0.95 & \cellcolor{mattePink15}0.32 & \cellcolor{mattePink15}0.43 \\
U+M     & \cellcolor{mattePink10}1.02 & \cellcolor{mattePink15}1.34 & \cellcolor{mattePink25}0.38 & \cellcolor{mattePink15}0.49 & \cellcolor{mattePink15}0.84 & \cellcolor{mattePink15}1.23 & \cellcolor{mattePink25}0.30 & \cellcolor{mattePink15}0.40 \\
U+XV    & \cellcolor{mattePink15}0.82 & \cellcolor{mattePink25}0.97 & \cellcolor{mattePink25}0.31 & \cellcolor{mattePink25}0.44 & \cellcolor{mattePink25}0.75 & \cellcolor{mattePink25}0.90 & \cellcolor{mattePink25}0.22 & \cellcolor{mattePink25}0.35 \\
U+EC    & \cellcolor{mattePink25}0.70 & \cellcolor{mattePink25}0.85 & \cellcolor{mattePink25}0.25 & \cellcolor{mattePink25}0.38 & \cellcolor{mattePink25}0.65 & \cellcolor{mattePink25}0.80 & \cellcolor{mattePink25}0.18 & \cellcolor{mattePink25}0.30 \\
U+m2v   & \cellcolor{mattePink10}0.98 & \cellcolor{mattePink15}1.30 & \cellcolor{mattePink5}0.46  & \cellcolor{mattePink10}0.54 & \cellcolor{mattePink10}0.94 & \cellcolor{mattePink15}1.17 & \cellcolor{mattePink10}0.36 & \cellcolor{mattePink15}0.44 \\
U+MT95  & \cellcolor{mattePink10}1.02 & \cellcolor{mattePink5}1.41  & \cellcolor{mattePink5}0.48  & \cellcolor{mattePink5}0.56  & \cellcolor{mattePink5}0.79  & \cellcolor{mattePink5}1.01  & \cellcolor{mattePink5}0.40  & \cellcolor{mattePink5}0.48 \\
U+MTP   & \cellcolor{mattePink5}1.08  & \cellcolor{mattePink5}1.56  & \cellcolor{mattePink5}0.52  & \cellcolor{mattePink5}0.58  & \cellcolor{mattePink5}0.90  & \cellcolor{mattePink5}1.07  & \cellcolor{mattePink5}0.44  & \cellcolor{mattePink5}0.50 \\
U+MT3M  & \cellcolor{mattePink10}1.02 & \cellcolor{mattePink5}1.37  & \cellcolor{mattePink5}0.50  & \cellcolor{mattePink5}0.56  & \cellcolor{mattePink5}0.82  & \cellcolor{mattePink5}0.92  & \cellcolor{mattePink5}0.48  & \cellcolor{mattePink5}0.55 \\
U+MTV0  & \cellcolor{mattePink5}1.06  & \cellcolor{mattePink5}1.50  & \cellcolor{mattePink5}0.50  & \cellcolor{mattePink5}0.56  & \cellcolor{mattePink5}0.88  & \cellcolor{mattePink5}1.01  & \cellcolor{mattePink5}0.42  & \cellcolor{mattePink5}0.48 \\
\midrule
W2+W    & \cellcolor{mattePink15}0.94 & \cellcolor{mattePink15}1.28 & \cellcolor{mattePink15}0.41 & \cellcolor{mattePink15}0.50 & \cellcolor{mattePink15}0.89 & \cellcolor{mattePink15}1.04 & \cellcolor{mattePink15}0.33 & \cellcolor{mattePink15}0.42 \\
W2+X    & \cellcolor{mattePink15}0.96 & \cellcolor{mattePink15}1.27 & \cellcolor{mattePink15}0.40 & \cellcolor{mattePink15}0.50 & \cellcolor{mattePink20}0.82 & \cellcolor{mattePink20}0.90 & \cellcolor{mattePink15}0.32 & \cellcolor{mattePink15}0.43 \\
W2+Wh   & \cellcolor{mattePink15}0.93 & \cellcolor{mattePink20}1.19 & \cellcolor{mattePink20}0.37 & \cellcolor{mattePink15}0.50 & \cellcolor{mattePink15}0.89 & \cellcolor{mattePink15}1.03 & \cellcolor{mattePink15}0.30 & \cellcolor{mattePink15}0.41 \\
W2+M    & \cellcolor{mattePink15}0.94 & \cellcolor{mattePink15}1.20 & \cellcolor{mattePink25}0.35 & \cellcolor{mattePink25}0.49 & \cellcolor{mattePink15}0.93 & \cellcolor{mattePink15}1.09 & \cellcolor{mattePink15}0.30 & \cellcolor{mattePink15}0.44 \\
W2+XV   & \cellcolor{mattePink25}0.83 & \cellcolor{mattePink25}0.96 & \cellcolor{mattePink25}0.32 & \cellcolor{mattePink25}0.44 & \cellcolor{mattePink25}0.74 & \cellcolor{mattePink25}0.89 & \cellcolor{mattePink25}0.22 & \cellcolor{mattePink25}0.35 \\
W2+EC   & \cellcolor{mattePink25}0.69 & \cellcolor{mattePink25}0.84 & \cellcolor{mattePink25}0.24 & \cellcolor{mattePink25}0.37 & \cellcolor{mattePink25}0.64 & \cellcolor{mattePink25}0.79 & \cellcolor{mattePink25}0.17 & \cellcolor{mattePink25}0.29 \\
W2+m2v  & \cellcolor{mattePink5}1.05  & \cellcolor{mattePink5}1.46  & \cellcolor{mattePink5}0.61  & \cellcolor{mattePink15}0.50 & \cellcolor{mattePink15}0.89 & \cellcolor{mattePink15}1.05 & \cellcolor{mattePink15}0.48 & \cellcolor{mattePink15}0.39 \\
W2+MT95 & \cellcolor{mattePink15}1.01 & \cellcolor{mattePink15}1.41 & \cellcolor{mattePink15}0.47 & \cellcolor{mattePink15}0.55 & \cellcolor{mattePink15}0.91 & \cellcolor{mattePink15}1.07 & \cellcolor{mattePink15}0.39 & \cellcolor{mattePink15}0.47 \\
W2+MTP  & \cellcolor{mattePink15}0.97 & \cellcolor{mattePink15}1.29 & \cellcolor{mattePink5}0.57  & \cellcolor{mattePink5}0.63  & \cellcolor{mattePink15}0.89 & \cellcolor{mattePink15}1.03 & \cellcolor{mattePink5}0.49  & \cellcolor{mattePink5}0.55 \\
W2+MT3M & \cellcolor{mattePink15}1.03 & \cellcolor{mattePink15}1.40 & \cellcolor{mattePink15}0.48 & \cellcolor{mattePink15}0.55 & \cellcolor{mattePink15}0.93 & \cellcolor{mattePink15}1.21 & \cellcolor{mattePink15}0.32 & \cellcolor{mattePink15}0.43 \\
W2+MTV0 & \cellcolor{mattePink15}1.01 & \cellcolor{mattePink15}1.36 & \cellcolor{mattePink15}0.48 & \cellcolor{mattePink15}0.55 & \cellcolor{mattePink15}0.91 & \cellcolor{mattePink15}1.08 & \cellcolor{mattePink15}0.40 & \cellcolor{mattePink15}0.47 \\
\midrule
W+X     & \cellcolor{mattePink15}0.96 & \cellcolor{mattePink15}1.25 & \cellcolor{mattePink15}0.39 & \cellcolor{mattePink15}0.50 & \cellcolor{mattePink20}0.82 & \cellcolor{mattePink20}1.19 & \cellcolor{mattePink15}0.31 & \cellcolor{mattePink15}0.42 \\
W+Wh    & \cellcolor{mattePink15}0.96 & \cellcolor{mattePink15}1.24 & \cellcolor{mattePink15}0.39 & \cellcolor{mattePink15}0.49 & \cellcolor{mattePink15}0.91 & \cellcolor{mattePink15}1.08 & \cellcolor{mattePink15}0.31 & \cellcolor{mattePink15}0.41 \\
W+M     & \cellcolor{mattePink25}0.88 & \cellcolor{mattePink25}1.07 & \cellcolor{mattePink15}0.40 & \cellcolor{mattePink15}0.50 & \cellcolor{mattePink15}0.85 & \cellcolor{mattePink15}1.00 & \cellcolor{mattePink15}0.32 & \cellcolor{mattePink15}0.42 \\
W+XV    & \cellcolor{mattePink25}0.82 & \cellcolor{mattePink25}0.95 & \cellcolor{mattePink25}0.31 & \cellcolor{mattePink25}0.43 & \cellcolor{mattePink25}0.73 & \cellcolor{mattePink25}0.88 & \cellcolor{mattePink25}0.21 & \cellcolor{mattePink25}0.34 \\
W+EC    & \cellcolor{mattePink25}0.68 & \cellcolor{mattePink25}0.83 & \cellcolor{mattePink25}0.23 & \cellcolor{mattePink25}0.36 & \cellcolor{mattePink25}0.63 & \cellcolor{mattePink25}0.78 & \cellcolor{mattePink25}0.16 & \cellcolor{mattePink25}0.28 \\
W+m2v   & \cellcolor{mattePink15}0.90 & \cellcolor{mattePink15}1.10 & \cellcolor{mattePink5}0.47  & \cellcolor{mattePink5}0.55  & \cellcolor{mattePink15}0.85 & \cellcolor{mattePink15}1.02 & \cellcolor{mattePink5}0.36  & \cellcolor{mattePink5}0.41 \\
W+MT95  & \cellcolor{mattePink15}0.94 & \cellcolor{mattePink15}1.20 & \cellcolor{mattePink5}0.64  & \cellcolor{mattePink5}0.53  & \cellcolor{mattePink15}0.91 & \cellcolor{mattePink15}1.08 & \cellcolor{mattePink5}0.61  & \cellcolor{mattePink5}0.47 \\
W+MTP   & \cellcolor{mattePink15}0.97 & \cellcolor{mattePink15}1.23 & \cellcolor{mattePink5}0.61  & \cellcolor{mattePink5}0.59  & \cellcolor{mattePink15}0.90 & \cellcolor{mattePink15}1.02 & \cellcolor{mattePink5}0.53  & \cellcolor{mattePink5}0.56 \\
W+MT3M  & \cellcolor{mattePink15}0.94 & \cellcolor{mattePink15}1.17 & \cellcolor{mattePink5}0.45  & \cellcolor{mattePink5}0.53  & \cellcolor{mattePink15}0.90 & \cellcolor{mattePink15}1.06 & \cellcolor{mattePink5}0.37  & \cellcolor{mattePink5}0.45 \\
W+MTV0  & \cellcolor{mattePink15}0.97 & \cellcolor{mattePink15}1.24 & \cellcolor{mattePink5}0.56  & \cellcolor{mattePink5}0.58  & \cellcolor{mattePink15}0.91 & \cellcolor{mattePink15}1.07 & \cellcolor{mattePink5}0.48  & \cellcolor{mattePink5}0.50 \\
\midrule
X+Wh    & \cellcolor{mattePink15}0.97 & \cellcolor{mattePink15}1.32 & \cellcolor{mattePink25}0.36 & \cellcolor{mattePink25}0.48 & \cellcolor{mattePink15}0.91 & \cellcolor{mattePink15}1.05 & \cellcolor{mattePink25}0.28 & \cellcolor{mattePink25}0.40 \\
X+M     & \cellcolor{mattePink15}0.94 & \cellcolor{mattePink15}1.20 & \cellcolor{mattePink25}0.37 & \cellcolor{mattePink25}0.50 & \cellcolor{mattePink15}0.88 & \cellcolor{mattePink15}1.06 & \cellcolor{mattePink25}0.29 & \cellcolor{mattePink25}0.42 \\
X+XV    & \cellcolor{mattePink25}0.81 & \cellcolor{mattePink25}0.94 & \cellcolor{mattePink25}0.30 & \cellcolor{mattePink25}0.42 & \cellcolor{mattePink25}0.72 & \cellcolor{mattePink25}0.87 & \cellcolor{mattePink25}0.21 & \cellcolor{mattePink25}0.34 \\
X+EC    & \cellcolor{mattePink25}0.67 & \cellcolor{mattePink25}0.82 & \cellcolor{mattePink25}0.22 & \cellcolor{mattePink25}0.35 & \cellcolor{mattePink25}0.62 & \cellcolor{mattePink25}0.77 & \cellcolor{mattePink25}0.15 & \cellcolor{mattePink25}0.27 \\
X+m2v   & \cellcolor{mattePink15}0.98 & \cellcolor{mattePink15}1.30 & \cellcolor{mattePink15}0.42 & \cellcolor{mattePink15}0.52 & \cellcolor{mattePink15}0.90 & \cellcolor{mattePink15}1.05 & \cellcolor{mattePink15}0.34 & \cellcolor{mattePink15}0.44 \\
X+MT95  & \cellcolor{mattePink5}1.06  & \cellcolor{mattePink5}1.54  & \cellcolor{mattePink15}0.48 & \cellcolor{mattePink15}0.55 & \cellcolor{mattePink15}0.91 & \cellcolor{mattePink15}1.08 & \cellcolor{mattePink15}0.40 & \cellcolor{mattePink15}0.47 \\
X+MTP   & \cellcolor{mattePink15}0.94 & \cellcolor{mattePink15}1.23 & \cellcolor{mattePink25}0.41 & \cellcolor{mattePink25}0.52 & \cellcolor{mattePink15}0.90 & \cellcolor{mattePink15}1.07 & \cellcolor{mattePink25}0.33 & \cellcolor{mattePink25}0.44 \\
X+MT3M  & \cellcolor{mattePink5}0.99  & \cellcolor{mattePink5}1.32  & \cellcolor{mattePink15}0.45 & \cellcolor{mattePink15}0.53 & \cellcolor{mattePink15}0.89 & \cellcolor{mattePink15}1.06 & \cellcolor{mattePink15}0.37 & \cellcolor{mattePink15}0.45 \\
X+MTV0  & \cellcolor{mattePink15}0.94 & \cellcolor{mattePink15}1.23 & \cellcolor{mattePink25}0.42 & \cellcolor{mattePink25}0.53 & \cellcolor{mattePink15}0.88 & \cellcolor{mattePink15}1.03 & \cellcolor{mattePink25}0.34 & \cellcolor{mattePink25}0.45 \\
\midrule
Wh+M    & \cellcolor{mattePink15}0.91 & \cellcolor{mattePink15}1.15 & \cellcolor{mattePink25}0.36 & \cellcolor{mattePink25}0.48 & \cellcolor{mattePink15}0.89 & \cellcolor{mattePink15}1.04 & \cellcolor{mattePink25}0.28 & \cellcolor{mattePink25}0.40 \\
Wh+XV   & \cellcolor{mattePink25}0.82 & \cellcolor{mattePink25}0.95 & \cellcolor{mattePink25}0.31 & \cellcolor{mattePink25}0.43 & \cellcolor{mattePink25}0.73 & \cellcolor{mattePink25}0.88 & \cellcolor{mattePink25}0.21 & \cellcolor{mattePink25}0.34 \\
Wh+EC   & \cellcolor{mattePink25}0.66 & \cellcolor{mattePink25}0.81 & \cellcolor{mattePink25}0.21 & \cellcolor{mattePink25}0.34 & \cellcolor{mattePink25}0.61 & \cellcolor{mattePink25}0.76 & \cellcolor{mattePink25}0.14 & \cellcolor{mattePink25}0.26 \\
Wh+m2v  & \cellcolor{mattePink15}0.97 & \cellcolor{mattePink15}1.29 & \cellcolor{mattePink25}0.44 & \cellcolor{mattePink25}0.53 & \cellcolor{mattePink15}0.89 & \cellcolor{mattePink15}1.05 & \cellcolor{mattePink25}0.36 & \cellcolor{mattePink25}0.45 \\
Wh+MT95 & \cellcolor{mattePink15}0.97 & \cellcolor{mattePink15}1.30 & \cellcolor{mattePink25}0.47 & \cellcolor{mattePink25}0.55 & \cellcolor{mattePink15}0.91 & \cellcolor{mattePink15}1.07 & \cellcolor{mattePink25}0.39 & \cellcolor{mattePink25}0.47 \\
Wh+MTP  & \cellcolor{mattePink15}1.01 & \cellcolor{mattePink15}1.37 & \cellcolor{mattePink25}0.45 & \cellcolor{mattePink25}0.54 & \cellcolor{mattePink15}0.91 & \cellcolor{mattePink15}1.08 & \cellcolor{mattePink25}0.37 & \cellcolor{mattePink25}0.46 \\
Wh+MT3M & \cellcolor{mattePink15}0.99 & \cellcolor{mattePink15}1.30 & \cellcolor{mattePink25}0.45 & \cellcolor{mattePink25}0.53 & \cellcolor{mattePink15}0.93 & \cellcolor{mattePink15}1.10 & \cellcolor{mattePink25}0.37 & \cellcolor{mattePink25}0.42 \\
Wh+MTV0 & \cellcolor{mattePink15}0.95 & \cellcolor{mattePink15}1.22 & \cellcolor{mattePink5}0.57  & \cellcolor{mattePink5}0.78  & \cellcolor{mattePink15}0.85 & \cellcolor{mattePink15}1.03 & \cellcolor{mattePink5}0.49  & \cellcolor{mattePink5}0.51 \\
\midrule
M+XV    & \cellcolor{mattePink25}0.80 & \cellcolor{mattePink25}0.93 & \cellcolor{mattePink25}0.29 & \cellcolor{mattePink25}0.41 & \cellcolor{mattePink25}0.71 & \cellcolor{mattePink25}0.86 & \cellcolor{mattePink25}0.20 & \cellcolor{mattePink25}0.33 \\
M+EC   & \cellcolor{mattePink25}0.65 & \cellcolor{mattePink25}0.80 & \cellcolor{mattePink25}0.20 & \cellcolor{mattePink25}0.33 & \cellcolor{mattePink25}0.60 & \cellcolor{mattePink25}0.75 & \cellcolor{mattePink25}0.13 & \cellcolor{mattePink25}0.25 \\
M+m2v   & \cellcolor{mattePink15}1.02 & \cellcolor{mattePink15}1.35 & \cellcolor{mattePink15}0.41 & \cellcolor{mattePink15}0.51 & \cellcolor{mattePink15}0.91 & \cellcolor{mattePink15}1.08 & \cellcolor{mattePink15}0.33 & \cellcolor{mattePink15}0.43 \\
M+MT95  & \cellcolor{mattePink15}0.95 & \cellcolor{mattePink15}1.22 & \cellcolor{mattePink15}0.41 & \cellcolor{mattePink15}0.52 & \cellcolor{mattePink15}0.88 & \cellcolor{mattePink15}1.02 & \cellcolor{mattePink15}0.33 & \cellcolor{mattePink15}0.44 \\
M+MTP   & \cellcolor{mattePink15}0.98 & \cellcolor{mattePink15}1.27 & \cellcolor{mattePink25}0.40 & \cellcolor{mattePink25}0.50 & \cellcolor{mattePink15}0.88 & \cellcolor{mattePink15}1.03 & \cellcolor{mattePink25}0.32 & \cellcolor{mattePink25}0.42 \\
M+MT3M  & \cellcolor{mattePink15}0.98 & \cellcolor{mattePink15}1.28 & \cellcolor{mattePink15}0.44 & \cellcolor{mattePink15}0.52 & \cellcolor{mattePink15}0.89 & \cellcolor{mattePink15}1.04 & \cellcolor{mattePink15}0.36 & \cellcolor{mattePink15}0.44 \\
M+MTV0  & \cellcolor{mattePink15}0.93 & \cellcolor{mattePink15}1.16 & \cellcolor{mattePink15}0.41 & \cellcolor{mattePink15}0.52 & \cellcolor{mattePink15}0.86 & \cellcolor{mattePink15}1.01 & \cellcolor{mattePink15}0.33 & \cellcolor{mattePink15}0.46 \\
\midrule
\textbf{XV+EC}   & \cellcolor{mattePink25}\textbf{0.56} & \cellcolor{mattePink25}\textbf{0.71} & \cellcolor{mattePink25}\textbf{0.14} & \cellcolor{mattePink25}\textbf{0.26} & \cellcolor{mattePink25}\textbf{0.49} & \cellcolor{mattePink25}\textbf{0.67} & \cellcolor{mattePink25}\textbf{0.09} & \cellcolor{mattePink25}\textbf{0.21} \\
XV+m2v  & \cellcolor{mattePink25}0.65 & \cellcolor{mattePink25}0.79 & \cellcolor{mattePink25}0.35 & \cellcolor{mattePink25}0.45 & \cellcolor{mattePink25}0.79 & \cellcolor{mattePink25}0.94 & \cellcolor{mattePink25}0.26 & \cellcolor{mattePink25}0.38 \\
XV+MT95 & \cellcolor{mattePink25}0.63 & \cellcolor{mattePink25}0.79 & \cellcolor{mattePink25}0.33 & \cellcolor{mattePink25}0.45 & \cellcolor{mattePink25}0.77 & \cellcolor{mattePink25}0.92 & \cellcolor{mattePink25}0.24 & \cellcolor{mattePink25}0.36 \\
XV+MTP  & \cellcolor{mattePink25}0.62 & \cellcolor{mattePink25}0.78 & \cellcolor{mattePink25}0.32 & \cellcolor{mattePink25}0.45 & \cellcolor{mattePink25}0.76 & \cellcolor{mattePink25}0.91 & \cellcolor{mattePink25}0.23 & \cellcolor{mattePink25}0.35 \\
XV+MT3M & \cellcolor{mattePink25}0.64 & \cellcolor{mattePink25}0.90 & \cellcolor{mattePink25}0.33 & \cellcolor{mattePink25}0.45 & \cellcolor{mattePink25}0.77 & \cellcolor{mattePink25}0.92 & \cellcolor{mattePink25}0.24 & \cellcolor{mattePink25}0.36 \\
XV+MTV0 & \cellcolor{mattePink25}0.63 & \cellcolor{mattePink25}0.88 & \cellcolor{mattePink25}0.34 & \cellcolor{mattePink25}0.45 & \cellcolor{mattePink25}0.76 & \cellcolor{mattePink25}0.90 & \cellcolor{mattePink25}0.23 & \cellcolor{mattePink25}0.35 \\
\midrule
EC+m2v  & \cellcolor{mattePink25}0.63 & \cellcolor{mattePink25}0.88 & \cellcolor{mattePink25}0.18 & \cellcolor{mattePink25}0.32 & \cellcolor{mattePink25}0.58 & \cellcolor{mattePink25}0.73 & \cellcolor{mattePink25}0.11 & \cellcolor{mattePink25}0.23 \\
EC+MT95 & \cellcolor{mattePink25}0.62 & \cellcolor{mattePink25}0.87 & \cellcolor{mattePink25}0.17 & \cellcolor{mattePink25}0.31 & \cellcolor{mattePink25}0.57 & \cellcolor{mattePink25}0.72 & \cellcolor{mattePink25}0.10 & \cellcolor{mattePink25}0.22 \\
EC+MTP  & \cellcolor{mattePink25}0.61 & \cellcolor{mattePink25}0.76 & \cellcolor{mattePink25}0.16 & \cellcolor{mattePink25}0.30 & \cellcolor{mattePink25}0.56 & \cellcolor{mattePink25}0.71 & \cellcolor{mattePink25}0.19 & \cellcolor{mattePink25}0.24 \\
EC+MT3M & \cellcolor{mattePink25}0.60 & \cellcolor{mattePink25}0.75 & \cellcolor{mattePink25}0.15 & \cellcolor{mattePink25}0.29 & \cellcolor{mattePink25}0.55 & \cellcolor{mattePink25}0.70 & \cellcolor{mattePink25}0.18 & \cellcolor{mattePink25}0.23 \\
EC+MTV0 & \cellcolor{mattePink25}0.59 & \cellcolor{mattePink25}0.84 & \cellcolor{mattePink25}0.17 & \cellcolor{mattePink25}0.28 & \cellcolor{mattePink25}0.54 & \cellcolor{mattePink25}0.69 & \cellcolor{mattePink25}0.17 & \cellcolor{mattePink25}0.29 \\
\midrule
m2v+MTP  & \cellcolor{mattePink15}0.93 & \cellcolor{mattePink15}1.18 & \cellcolor{mattePink5}0.74  & \cellcolor{mattePink5}2.02  & \cellcolor{mattePink15}0.90 & \cellcolor{mattePink15}1.06 & \cellcolor{mattePink15}0.62 & \cellcolor{mattePink15}0.65 \\
m2v+MT3M & \cellcolor{mattePink15}0.96 & \cellcolor{mattePink15}1.53 & \cellcolor{mattePink5}0.72  & \cellcolor{mattePink5}3.62  & \cellcolor{mattePink15}0.89 & \cellcolor{mattePink15}1.05 & \cellcolor{mattePink15}0.62 & \cellcolor{mattePink15}0.62 \\
\midrule
MT95+MT3M & \cellcolor{mattePink15}0.95 & \cellcolor{mattePink15}1.20 & \cellcolor{mattePink5}0.84  & \cellcolor{mattePink5}0.84  & \cellcolor{mattePink15}0.86 & \cellcolor{mattePink15}1.02 & \cellcolor{mattePink5}0.59  & \cellcolor{mattePink5}0.63 \\
MT95+MTV0 & \cellcolor{mattePink15}0.94 & \cellcolor{mattePink15}1.21 & \cellcolor{mattePink5}0.66  & \cellcolor{mattePink5}1.50  & \cellcolor{mattePink15}0.88 & \cellcolor{mattePink15}1.03 & \cellcolor{mattePink5}0.59  & \cellcolor{mattePink5}0.61 \\
MT95+MTV0 & \cellcolor{mattePink5}1.05  & \cellcolor{mattePink5}1.53  & \cellcolor{mattePink5}0.66  & \cellcolor{mattePink5}0.66  & \cellcolor{mattePink15}0.88 & \cellcolor{mattePink15}1.04 & \cellcolor{mattePink5}0.60  & \cellcolor{mattePink5}0.60 \\
\midrule
MTP+MT3M  & \cellcolor{mattePink15}0.95 & \cellcolor{mattePink15}1.25 & \cellcolor{mattePink5}0.63  & \cellcolor{mattePink5}0.74  & \cellcolor{mattePink15}0.88 & \cellcolor{mattePink15}1.03 & \cellcolor{mattePink5}0.59  & \cellcolor{mattePink5}0.62 \\
MTP+MTV0  & \cellcolor{mattePink15}0.91 & \cellcolor{mattePink15}1.14 & \cellcolor{mattePink5}0.64  & \cellcolor{mattePink5}0.64  & \cellcolor{mattePink15}0.86 & \cellcolor{mattePink15}1.01 & \cellcolor{mattePink5}0.59  & \cellcolor{mattePink5}0.64 \\
\midrule
MT3M+MTV0 & \cellcolor{mattePink15}0.96 & \cellcolor{mattePink15}1.22 & \cellcolor{mattePink5}0.60  & \cellcolor{mattePink5}0.62  & \cellcolor{mattePink15}0.85 & \cellcolor{mattePink15}1.02 & \cellcolor{mattePink5}0.57  & \cellcolor{mattePink5}0.60 \\
\bottomrule
\end{tabular}
\caption{Performance Scores of different PTM combinations; TO1, TM stands for test-other1 and test-main}
\label{tab:fusion}
\end{table}

\vspace{-0.3cm}

\subsection{Experimental Results}
We use Mean Squared Error (MSE) and Mean Absolute Error (MAE) as evaluation metrics \cite{tang2024singmos}. Table \ref{tab:single} presents the evaluation results of models trained with individual PTM representations for SingMOS prediction. We present the results on test-other1 and test-main testing set. Our results demonstrate that speaker recognition SPTMs consistently outperform other SPTMs as well as MPTMs with the lowest MAE and MSE with both FCN and CNN downstreams in both test sets: test-other1 and test-main. This validates \textit{our hypothesis that speaker recognition SPTMs can significantly improve SingMOS prediction. Their pre-training enables them to capture essential vocal attributes—pitch, tone, intensity, and rhythm more effectively than other SPTMs and MPTMs that are crucial for evaluating generated singing voice quality.} Among the speaker recognition SPTMs, there is no clear champion with both attaining mixed performance. Such mixed results are also obtained for both monolingual and multilingual SPTMs thus pointing towards the relevance on downstream data distribution on the evaluation scores. Within monolingual SPTMs, we see that Unispeech-SAT obtained the lowest performance with CNN in MSE (0.90 \%) while Wav2vec2 showing the best result with CNN in MSE (0.77\%). The low performance of monolingual SPTMs can also be due to the language variability during their pre-training and the downstream task of SingMOS prediction. As the selected monolingual SPTMs are trained on english and the dataset considered in our study, consists of Japanese and Chinese voice samples. \textit{However, according to this logic, as the multilingual SPTMs should have shown good performance on SingMOS prediction as these multilingual SPTMs have been exposed to Chinese and Japanese during their multilingual pre-training. However, that's not the case and speaker recognition SPTMs attains the topmost position. This point further emphasizes on our proposed hypothesis and its utility for SingMOS prediction.} Multilingual SPTMs shows comparable and sometimes better performance than the monolingual SPTMs across both the test sets. MPTMs in both the test sets attains comparable or lower performance than its SPTMs counterparts. This can be its inability to capture the important vocal characteristics from the synthesized singing voices for SingMOS prediction. Overall, the CNN models showed better performance than the FCN models. \par
Table \ref{tab:fusion} presents the results of combinations of different PTMs. We use concatenation-based fusion as the baseline fusion technique. We keep the same modeling as given in Figure \ref{fig:archi} except we remove the gate mechanism and BD loss. We also keep the training details as \textbf{\texttt{BATCH}} for fair comparison. We observe that fusion through \textbf{\texttt{BATCH}} with various combinations of PTMs attains relatively better performance than concatentation-based fusion technique. These results shows the effective fusion capability of \textbf{\texttt{BATCH}}. With \textbf{\texttt{BATCH}} through the combination of speaker recognition SPTMs x-vector and ECAPA, we attained the topmost performance across the possible combinations of PTMs encompassing diverse PTMs ranging from monolingual, multilingual as well as MPTMs. The speaker recognition SPTMs had already shown its capability for SingMOS prediction individually in Table \ref{tab:single}, but, when in unison these SPTMs performs even more better and thus bringing their complementary strength. Overall, we generally observe that fusion of any other SPTMs and MPTMs with speaker recognition SPTMs shows better performance than the individual results of those particular SPTMs and MPTMs. However, we observe that certain combinations doesn't always improve performance and this can be due to the conflicting behavior representation space from the PTMs. For example, combination of Unispeech-SAT and Wav2vec2 through \textbf{\texttt{BATCH}} on test-other1 reports MAE of 0.91\% which is worst than their individual MAE scores of 0.79\% and 0.77\% with CNN for Unispeech-SAT and Wav2vec2, thus showing their conflicting behavior. Furthermore, we can see that experiments with concatenation-based fusion doesn't show much improvement in comparison to respective individual PTMs results whose fusion is considered. In contrast, \textbf{\texttt{BATCH}} with different the PTMs combinations shows better or comparable performance with individual PTMs in both test sets. 

\noindent \textbf{Comparison to Methods used in Previous Works}: We present a comparison of our best approaches with methods used in previous SOTA works \cite{tang2024singmos, tang2024exploration}. These studies showed that they obtain top performances with SPTMs such as XLS-R, Wav2vec2, WavLM. Here, we perform experiments with these SPTMs used in previous studies and show that speaker recognition SPTMs are the best for SingMOS prediction amongst various SOTA SPTMs including XLS-R, Wav2vec2, WavLM as well as MPTMs. Further, their fusion through \textbf{\texttt{BATCH}} reports much more improved performance and thus setting SOTA for SingMOS prediction. 

\section{Conclusion}
In this study, we showed that the speaker recognition SPTMs such as x-vector, ECAPA are the most effective for SingMOS prediction. We validated it by presenting comprehensive comparative analysis of various SOTA PTMs including monolingual, multilingual, speaker recognition SPTMs as well as MPTMs. Further, we proposed, \textbf{\texttt{BATCH}} for effective fusion of PTMs and we show that fusion of speaker recognition SPTMs brings the best out of them by showing complementary behavior for more improved SingMOS prediction. Our study highlights the overlooked potential of speaker recognition PTMs for SingMOS prediction, providing a strong foundation for future research in this domain and will act as a reference for future researchers exploring PTMs for SingMOS prediction.

\bibliographystyle{IEEEtran}
\bibliography{main}

\end{document}